# PLD-Based Reconfigurable Controllers for Feedback Systems

V.F. Gubarev, L.M. Yakovleva, N.N. Aksenov, Space Research Institute, Kiev, Ukraine
A.V. Palagin, V.N. Opanasenko, V.G. Sakharin, Institute of Cybernetics, Kiev, Ukraine

Abstract

Recently, interest has grown for application of reconfigurable devices in robust and adaptive control systems. The main advantage of such devices is that its structure is not fixed and may be varied depending on the currently used control algorithm. In this paper a new PC-based reconfigurable microcontrollers (RMC) are offered for Experimental Physics control systems. Programmable logic devices (PLD) XILINX Inc. of a type FPGA are used in RMC. Reconfigurable devices are connected to PC as the coprocessor through the system bus (for example, PCI) forming the system with programmable architecture as a result. Each control algorithm is realized as a file of a PLD-configuration. The set of model- based control algorithms make up the functional library. Model selection procedure on the base of control relevant identification is a part of united control system design. Methodology of the joint and iterative identification and control is proposed which well suited for the realization in RMC and convenient for users. Control design is arranged as follows. First the identification experiment is carried out and approximate model is reconstructed using special identification algorithm and model structure in normalized form. Then select from the library the appropriate liner or nonlinear parametrized control law. After that the parametric class of controllers which guarantee stability of the closed loop system is automatically determined. The best robust controller in given class is obtained by parameter tuning procedure, which give opportunity to impact performance of the transient processes in closed system without loosing asymptotic stability.

## 1 INTRODUCTION

The microcontrollers are applied in modern control system's of technological processes, instrumentation, robots - manipulators, physical and other systems of real time, intended for to execution of the large sizes of calculations. They are realized, as a rule, or as sets of modules, which are built in the computer or as standalone control system's of the equipment, technological processes etc. For record of the control algorithm to the memory can be used ROM (the program is written at manufacture of the microcontrollers or with the help of the programmator) and RAM allowing to rewrite the program practically an infinite number of times.

Typically, the Experimental Physics and Industrial Control System (EPICS) is available on many platforms: VME -, VXI- and PC-based input output controllers. Advantage PC-based controllers is that there is only one software model for all PC motherboard, whereas there are typically many different software models for each of the many different VME-based of single board computers. In this paper new PC-based class of reconfigurable microcontrollers (RMC), which architecture varies with a program way is considered. The difference between reconfigurable and programmable microcontrollers is that their structure is not fixed and varies depending on the control algorithm. Thus the computational efficiency of algorithm is bounded by hardware only that is essential for real time systems.

PLD (programmable logic devices) Xilinx Corporation of a type FPGA (Field Programmable Gate Array) are used as element base for RMC mainly [1]. Essential advantage of PLD is their universality and possibility of fast program customization on the given control algorithm.

## 2 REALIZATION OF SET-VALUED CONTROL SYSTEM

Offered RMC is a new approach to construction of control system of experimental physical plants. This approach allows for multiple implementation to change both the parameters and structure, laws and strategies of control. It is especially important for physical plants when the conditions of experiment, targets and the tasks are varied. It is desirable for the experimenters and operators working with the system, without substitution of the equipment, to change algorithms and target of control, not breaking thus of stability and other important dynamic parameters of the system.

The system consists of standard hardware and reconfigurable coprocessors on base FPGA type PLD and set of programs realizing algorithms of iterated synthesis of control with identification of an approximating model. The creation of the well-structured library of algorithms and structural implementations, appropriate to them stored as files of configurations, is supposed. The task of synthesis in this case consists in a choice of optimal pair (algorithm-structure) for concrete source experimental data. Thus, the task of optimal synthesis is reduced to the task of an optimal choice on previously generated (and constantly extended) set of solutions. Distinctive feature of the offered controller is the possibility of creation of two-level nonlinear stabilization system, in which the lower automatic level provides stability and robustness of the close-looped system, and the top level allows to the engineer flexibly to operate quality of regulation.

The information for synthesis is measuring responses of the system on inputs. It is supposed also that the input and output information can be entered in PC using converters (analogue / digit and digit / analogue).

## 3 RECONFIGURABLE STRUCTURES

Usually structures of reconfigurable devices contain one or several chips PLD, memory for storage a configuration files (PROM or FLASH-memory), port of testing / debugging JTAG IEEE Std. 1149.1, SRAM or DRAM-memory for caching data, plug and controller for connection to buses ISA, PCI (for modules-coprocessors), plugs for connection of peripherals etc. The configuration file can be written down in ROM and automatically (at power up) to boot in a chip PLD. In the other mode a configuration file loads in a chip PLD from RAM of the computer.

The process of designing of PLD-based devices is completely supported by tools CAD, for example, Foundation Series of the Xilinx Corporation. Structure Foundation Series includes the library of parametric modules [2], which supports functional solutions in a range from small blocks (adders, registers, multipliers) up to larger blocks (correlators, DFT- 32, 64 and 128 points, FFT - 1024 points). The possibility of reconfiguration allows to make modifications of a working product or to use this product for implementation of the various control algorithms taking into account the specific features of hardware. The architecture of chips of series Virtex is completely compatible with technology IRL (Internet Reconfigurable Logic). Technology IRL gives the possibility of remote dynamic reconfiguration through the Internet [3], and also share development of standalone system or chips through Internet by various workgroups.

The abstract architecture RMC can be described as follows:

$$D = \langle G, B_i, F \rangle,$$

where: $G = \{G_i\}$ is set of control objects ($i = 1 \div n$);
$B_i = \{B_{ij}\}$ is set of the control algorithms, $B_i : X_{ij} \Rightarrow Y_{ij}$ where $\{X_{ij}\}$ is the set of input signals and $\{Y_{ij}\}$ is the set of output signals for $i$ - th object ($j = 1 \div m$);
$F = \{F_\mu\}$ is a configuration files set ($\mu = 1 \div k, k = n \times m$), defining structure of algorithms $B_{ij}$ for control objects $G_i$. After loading a configuration file in FPGA the structure of the device for implementation of appropriate algorithm include the structure of operational and internal control device will be generated.

## 4 IDENTIFICATION FOR CONTROL

It is offered to realize the model-based control system. The structure of a model can be known or to be determined in identification experiment The uniform methodology is developed which provides identification well suited for set of the control algorithms, i.e. identification and control are completely coordinated with each other. As the target is put to change by a desirable manner initial dynamic characteristics of the controlled object, this process can be carried out by an interactive way with restoring in identification experiment only of that part of a model, which we are going to change with the help of feedback. For an obtained submodel in the class of results guaranteeing stability and other important properties of transients, for example, robustness, the interactive way selects best feedback from the point of view of the experimenter. Thus can vary both parameters, and structure of feedback. If the result does not satisfy the experimenter, then identification of the system is repeated, but already for the close-looped system. Thus a submodel is identified again and parameters will be improved with the help of new feedback. The iterative process can proceed while obtaining desirable result.

For such iterative procedure of control synthesis was converging to desirable result, in the developed methodology the following two ways are used. The first one concerns to a choice of structure of an identifying model, which accepts splitting it on standalone blocks or subsystems. Usually it is canonical or normal forms of representation of the equations describing the system dynamic. The equivalent representation can be written down for discrete analogue of this dynamic system.

In the close-looped system the eigenvalues of the matrix corresponding to only subsystem are varied. All other roots of the characteristic equation are saved. Therefore second way of the developed methodology consists in estimation of considered submodel on base output of a current state. The estimation method advanced in [4] with use of observation results on a finite sliding interval is applied for this purpose.

## 5 LIBRARY OF CONTROL ALGORITHMS

Set of the admissible feedback laws, which can be realized in the control system, in a general case is written in matrix form as follows:

$$U = C^{(i)}(t,x) \cdot x, \ i = \overline{1, N},$$

where $x$ is the state vector of the dynamic system and set of matrices represents the some classes of the control laws.

Each class of appropriate controllers has the structure, in which the concrete implementation is satisfied by the defined set of parameters. For example, the class $C^{(0)}$ will generate a subset of linear regulators, which satisfy to a matrix $C^{(0)}$ with the constant elements. Other classes correspond to the nonlinear laws of regulation of various structures.

The distinctive feature of the offered methodology of share synthesis of control and identification is the iterative procedure with submodels of the low order used on each step and the nonlinear feedback laws. The nonlinear laws have more wide potentialities then linear ones, but their synthesis by known methods was rather bulky. It is proposed the formalized procedure of construction of set of nonlinear regulators on the base of Lyapunov functions and Sylvestr inequalities allowing to obtain areas of acceptable values of parameters, to select a subset of robust regulators, to optimize parameters of a regulator with the regard for the system of contradictory criterions. The structures of nonlinear regulators with tuning parameters are generated which can be changed over a wide range, effecting on quality of transients, not breaking thus of stability of the close-looped system.

## 6 CONTROL SYSTEM STRUCTURE

The library of configuration files (LCF) can settle down as in internal memory RMC, and, at the large size, in RAM PC. The swapping of configuration files from memory PC in RMC is made if necessary. The realization of the control system with PLD-based controller assumes library-building configuration files for hardware implementation of set of the control algorithms. Further on the base of this system and generated library LCF the experimenter in an interactive mode carries out the process of identification and synthesis of control. The experimenter on the display PC directly observes a system response to anyone controls and determines most appropriate feedback.

The external digital ports allow to carry out one- and multichannel input/output data processing Besides RMC can be used autonomously, irrespective of the computer. Additional logical resources in this case are released at the expense of absence of the controller of the external bus.

## 7 MANUSCRIPTS

Customization of structure for realization of chosen algorithm (or its fragment) and it implementation in a chip at a gate level allow to increase speed of the system in comparison with program solutions and to execute set of various algorithms with speed of specialized hardware.

The work of the experimenter with the offered control system does not require a special knowledge of the control synthesis and all operation is reduced to simple following of the instructions and recommendations applied to the system.